# Space Charge Effects

*M. Ferrario, M. Migliorati, and L. Palumbo*
INFN-LNF and University of Rome 'La Sapienza'

**Abstract**
The space charge forces are those generated directly by the charge distribution, with the inclusion of the image charges and currents due to the interaction of the beam with a perfectly conducting smooth pipe. Space charge forces are responsible for several unwanted phenomena related to beam dynamics, such as energy loss, shift of the synchronous phase and frequency, shift of the betatron frequencies, and instabilities. We will discuss in this lecture the main feature of space charge effects in high-energy storage rings as well as in low-energy linacs and transport lines.

## 1    Introduction

Charged particles moving in a linear or circular accelerator are guided, confined, and accelerated by external electromagnetic (e.m.) fields. In particular, the electric field in RF cavities is responsible for the acceleration, while the magnetic fields guide and focus the particles: the bending magnets are used to guide (i) the charges on the reference trajectory (orbit), (ii) the solenoids or quadrupoles with respect to the transverse confinement, and (iii) the sextupoles for the chromaticity correction.

The particle motion is governed by the Lorentz force via the following equation:

$$\frac{\mathrm{d}(m_0 \gamma \vec{v})}{\mathrm{d}t} = \vec{F}^{\mathrm{ext}} = e(\vec{E} + \vec{v} \times \vec{B}), \tag{1}$$

where $m_0$ is the rest mass, $\gamma$ is the relativistic factor, and $\vec{v}$ is the particle velocity. Using Eq. (1), we can in principle calculate the trajectory of the charge moving through any e.m. field.

The external forces $\vec{F}^{\mathrm{ext}}$ used for the beam transport and expressed by Eq. (1) do not depend on the beam current. In a real accelerator, however, there is another important source of e.m. fields to be considered, i.e. the beam itself, which circulates inside the pipe producing additional e.m. fields called 'self-fields'. These fields, which depend on the intensity of the beam current and the charge distribution, perturb the external guiding fields.

The self-fields are responsible for several unwanted phenomena related to beam dynamics, such as energy loss, shift of the synchronous phase and frequency, shift of the betatron frequencies, and instabilities. It is customary to divide the study of self-fields into space charge fields and wakefields. The space charge forces are those generated directly by the charge distribution, with the inclusion of the image charges and currents due to the interaction of the beam with a perfectly conducting smooth pipe [1]. The wakefields are instead produced by the finite conductivity of the walls and by all geometric variation of the beam pipe (such as resonant devices and transitions of the beam pipe). A reference paper on wakefields can be found in Ref. [2].

In this lecture we shall discuss only space charge effects, which are actually a particular case of Coulomb interactions in a multiparticle system. The net effect of the Coulomb interaction in a multiparticle system can be classified into two regimes [3]:

i) the *collisional regime*, dominated by binary collisions caused by close particle encounters, i.e. single-particle scattering,

ii) the *collective regime* or *space charge regime*, dominated by the self-field produced by the particle distribution, which varies appreciably only over large distances compared to the average separation of the particles.

The collisional part of the total interaction force arises when a particle is scattered by its immediate neighbours. This force will cause small random displacements of the particle's trajectory and statistical fluctuations in the particle distribution as a whole, leading for example to intra-beam scattering effects in high-energy storage rings [4] (see also the Touschek effect [5]). On the other hand, space charge forces lead to collective behaviour of the beam, driving for example envelope oscillations, emittance, and energy spread growth [6].

A measure of the relative importance of collisional versus collective effects in a beam is the Debye length, $\lambda_D = \sqrt{\varepsilon_0 \gamma^2 k_B T / e^2 n}$, where $n$ is the particle density, and the transverse beam temperature $T$ is defined as $k_B T = \gamma m_0 \langle v_\perp^2 \rangle$, where $k_B$ is the Boltzmann constant [3]. If a test charge is placed inside the beam, the excess electric potential $\Phi_D$ set up by this charge is effectively screened off in a distance $\lambda_D$ by charge redistribution in the beam as $\Phi_D(\vec{r}) = \frac{C}{r} e^{-r/\lambda_D}$. This effect is well known from plasma physics as *Debye shielding* [7]. A charged particle beam in a particle accelerator can be viewed as a non-neutral plasma [8] in which the smooth focusing channel replaces the restoring force produced by ions in a neutral plasma; see Fig. 1. As in a neutral plasma, the collective behaviour of the beam can be observed on length scales much larger than the Debye length. It follows that if the Debye length is small compared to the beam radius, collective effects due to the self-fields of the beam will play a dominant role in driving the beam dynamics with respect to binary collisions.

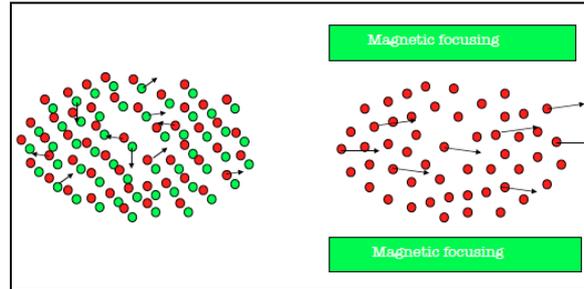

**Fig. 1:** The restoring force produced by the ions (green dots) in a neutral plasma can be replaced by a smooth focusing channel for charged particle beam (non-neutral plasma) in a particle accelerator.

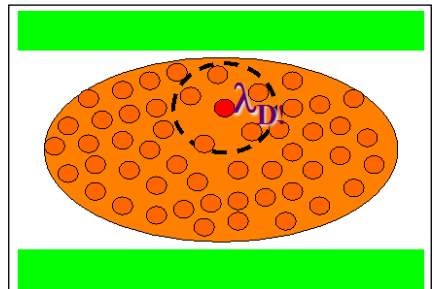

**Fig. 2:** Representation of the Debye sphere surrounding a test particle (red) in a beam dominated by space charge smooth fields (uniform orange).

Smooth functions for the charge and field distributions can be used, as in Section 3, as long as the Debye length remains large compared to the interparticle distance $d = n^{-1/3}$, that is as long as the number $N_D$ of particles within a Debye sphere of radius $\lambda_D$ remains large ($s = v_z t \gg 1$). A typical particle is actually scattered by all the other particles within its Debye sphere, but the large number of random interactions very rarely causes any sudden change in its motion (weakly coupled plasma), and mainly contributes to driving the beam toward thermal equilibrium [3].

The smoothed space charge forces acting on a particle can therefore be treated like an external force, and can be separated into linear and non-linear terms as a function of displacement from the beam axis. The linear space charge term typically defocuses the beam and leads to an increase in beam size. The non-linear space charge term increases the rms emittance by distorting the phase-space distribution. We shall see in Section 6 that the linear component of the space charge field can also induce reversible emittance growth in a bunched beam when longitudinal/transverse correlations along the bunch are taken into account.

Note that the Debye length increases with particle energy $\gamma$, so that at sufficiently high energy a transition from the space charge to the collisional regime may occur.

## 2    Self-fields and equations of motion

### 2.1    Betatron motion

Before dealing with the self-induced forces produced by the space charge and their effect on the beam dynamics in a circular accelerator, we briefly review the transverse equations of motion [9]. In order to simplify our study, let us consider a perfectly circular accelerator with radius $\rho_x$ and obtain the transverse single-particle equation of motion in the linear regime.

If in the particle equation of motion given by Eq. (1) we include the self-induced forces, we have

$$\frac{\mathrm{d}(m_0 \gamma \vec{v})}{\mathrm{d}t} = \vec{F}^{\mathrm{ext}}(\vec{r}) + \vec{F}^{\mathrm{self}}(\vec{r}). \tag{2}$$

By considering a constant energy $\gamma$, this becomes

$$\frac{\mathrm{d}\vec{v}}{\mathrm{d}t} = \frac{\vec{F}^{\mathrm{ext}}(\vec{r}) + \vec{F}^{\mathrm{self}}(\vec{r})}{m_0 \gamma}. \tag{3}$$

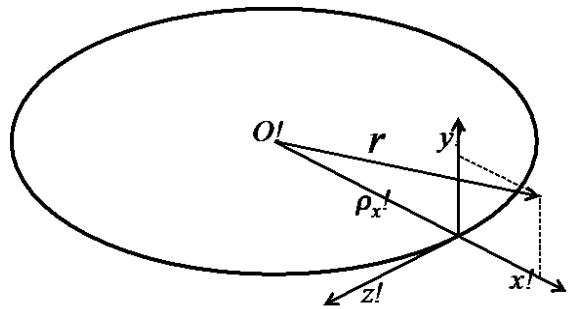

**Fig. 3:** Coordinate system for a charge in a circular accelerator

According to the coordinate system of Fig. 3, indicating the charge position by $\vec{r}$ and the transverse displacements with respect to the reference trajectory by $x$ and $y$, we write

$$\vec{r} = (\rho_x + x)\hat{e}_x + y\hat{e}_y. \tag{4}$$

Since the unit vector $\hat{e}_x$ rotates with angular frequency $\omega_0$ (clockwise in the figure), its time derivative is $\omega_0 \hat{e}_z$, so that the velocity is given by

$$\vec{v} = \frac{d\vec{r}}{dt} = \dot{x}\hat{e}_x + \dot{y}\hat{e}_y + \omega_0(\rho_x + x)\hat{e}_z, \tag{5}$$

and the acceleration is given by

$$\vec{a} = \left[\ddot{x} - \omega_0^2(\rho_x + x)\right]\hat{e}_x + \ddot{y}\hat{e}_y + \left[\dot{\omega}_0(\rho_x + x) + 2\omega_0\dot{x}\right]\hat{e}_z. \tag{6}$$

In the above equations, the dots refer to the derivative with respect to time.

If we consider the motion along $x$, we have

$$\ddot{x} - \omega_0^2(\rho_x + x) = \frac{1}{m_0\gamma}\left(F_x^{\text{ext}} + F_x^{\text{self}}\right). \tag{7}$$

Instead of using the time $t$ as variable, we consider the azimuthal position $s = v_z t$, so that the acceleration along $x$ becomes

$$\ddot{x} = \frac{d^2 x}{dt^2} = v_z^2 \frac{d^2 x}{ds^2} = v_z^2 x'' = \omega_0^2(\rho_x + x)^2 x'', \tag{8}$$

for which we have also used Eq. (5). Using Eq. (8), the differential equation of motion Eq. (7) can be written as

$$x'' - \frac{1}{\rho_x + x} = \frac{1}{m_0 v_z^2 \gamma}\left(F_x^{\text{ext}} + F_x^{\text{self}}\right). \tag{9}$$

We now assume small transverse displacements $x$ with respect to the machine radius $\rho_x$, so that we can linearly expand Eq. (9) as follows:

$$x'' - \frac{1}{\rho_x} + \frac{1}{\rho_x^2}x = \frac{1}{m_0 v_z^2 \gamma}\left(F_x^{\text{ext}} + F_x^{\text{self}}\right). \tag{10}$$

In addition, we have that the external force is due to the magnetic guiding fields. We assume that we have only dipoles and quadrupoles, or equivalently we expand the external guiding fields in a Taylor series up to the quadrupole component, i.e.

$$-F_x^{\text{ext}} = qv_z B_y = qv_z B_{y0} + qv_z \left(\frac{\partial B_y}{\partial x}\right)_0 x + \dots, \tag{11}$$

and recognize that the dipolar magnetic field $B_{y0}$ is responsible for the circular motion along the reference trajectory of radius $\rho_x$ according to

$$qv_z B_{y0} = \frac{m_0 \gamma v_z^2}{\rho_x}. \tag{12}$$

Finally, we obtain

$$x'' + \left[\frac{1}{\rho_x^2} + \frac{q}{m_0 v_z \gamma}\left(\frac{\partial B_y}{\partial x}\right)\right]x = \frac{1}{m_0 v_z^2 \gamma} F_x^{\text{self}}, \tag{13}$$

which can also be written as

$$x'' + \left[\frac{1}{\rho_x^2} - k\right]x = \frac{1}{m_0 v_z^2 \gamma} F_x^{\text{self}}, \quad (14)$$

where we have introduced the normalized gradient

$$k = \frac{g}{p/q} = -\frac{q}{m_0 v_z \gamma}\left(\frac{\partial B_y}{\partial x}\right), \quad (15)$$

Where $g$ is the quadrupole gradient [T/m] and $p$ is the charge momentum.

Equation (14) is not exactly correct, because both the curvature radius and the normalized gradient depend on the azimuthal position $s$. By using the focusing constant $K_x(s)$, we should write

$$x''(s) + K_x(s)x(s) = \frac{1}{m_0 v_z^2 \gamma} F_x^{\text{self}}(x,s). \quad (16)$$

In the absence of self-fields, the solution of the free equation (Hill's equation) yields the well-known betatron oscillations:

$$x(s) = a_x \sqrt{\beta_x(s)} \cos[\mu_x(s) - \varphi_x], \quad (17)$$

where $a_x$ and $\varphi_x$ depend on the initial conditions, and

$$\begin{aligned}&\frac{1}{2}\beta_x \beta_x'' - \frac{1}{4}\beta_x'^2 + K_x(s)\beta_x^2 = 1,\\ &\mu_x'(s) = 1/\beta_x(s),\\ &Q_x = \frac{\omega_x}{\omega_0} = \frac{1}{2\pi}\int_0^L \frac{ds'}{\beta_x(s')},\end{aligned} \quad (18)$$

where $Q_x$ is the betatron tune.

In the analysis of the motion in the presence of the self-induced fields, however, we adopt a simplified model in which particles execute simple harmonic oscillations around the reference trajectory. This is equivalent to having the focusing term $K_x$ constant around the machine. Although this case is never fulfilled in a real accelerator, it provides a reliable model for the description of the beam instabilities. Under this approximation Eq. (16) becomes

$$x''(s) + K_x x(s) = \frac{1}{m_0 v_z^2 \gamma} F_x^{\text{self}}(x,s), \quad (19)$$

which is a linear differential equation. The homogeneous solution is given by

$$x(s) = A_x \cos\left[\sqrt{K_x} s - \varphi_x\right], \quad (20)$$

where, using the notation of Eq. (18), we have

$$a_x\sqrt{\beta_x} = A_x,$$

$$\beta_x = \frac{1}{\mu_x'} = \frac{1}{\sqrt{K_x}},$$

$$\mu_x(s) = \sqrt{K_x}\, s,$$ (21)

$$Q_x = \frac{1}{2\pi}\int_0^L \frac{ds'}{\beta_x(s')} = \frac{L}{2\pi\beta_x} = \rho_x\sqrt{K_x} \Rightarrow K_x = \left(\frac{Q_x}{\rho_x}\right)^2.$$

The differential equation of motion (19) then becomes

$$x''(s) + \left(\frac{Q_x}{\rho_x}\right)^2 x(s) = \frac{1}{m_0 v_z^2 \gamma} F_x^{\text{self}}(x,s).$$ (22)

An analogous equation of motion can be written for the vertical plane:

$$y''(s) + \left(\frac{Q_y}{\rho_x}\right)^2 y(s) = \frac{1}{m_0 v_z^2 \gamma} F_y^{\text{self}}(y,s).$$ (23)

Equations (22) and (23) represent our starting point in the study of the effects of the self-induced fields on the betatron oscillations. Before analysing these forces, let us write the analogous equation for the longitudinal dynamics.

## 2.2 Synchrotron motion

In the longitudinal case, the motion is governed by the RF voltage, which we write as

$$V(t) = \hat{V}\sin[\omega_{\text{RF}} t + \varphi_s],$$ (24)

where $\varphi_s$ is the synchronous phase. In the linear approximation, and in absence of the self-induced forces, the equation of motion is that of a simple harmonic oscillator,

$$z'' + \left(\frac{Q_z}{\rho_x}\right)^2 z = 0,$$ (25)

with particles oscillating around the synchronous phase $\varphi_s$, with a synchronous tune given by

$$Q_z = \frac{\omega_z}{\omega_0} = \sqrt{\frac{qh\eta\hat{V}\cos\varphi_s}{2\pi\beta^2 E_0}},$$ (26)

where $h$ is the harmonic number, $E_0$ is the machine nominal energy, and

$$\eta = \frac{1}{\gamma^2} - \alpha_c.$$ (27)

The slippage factor $\eta$ accounts for the increase of the speed with energy $\left(1/\gamma^2\right)$ and the length of the real orbit due to the dispersion ($\alpha_c$).

The interaction of the charge with the surroundings may induce longitudinal e.m. forces, which have to be included in the equation of motion:

$$z'' + \left(\frac{Q_z}{\rho_x}\right)^2 z = \frac{\eta\, F_z^{self}(s)}{m_0 v_z^2 \gamma}\,. \qquad (28)$$

In the linear approximation, the longitudinal force produces a shift of both the synchronous phase and the synchronous tune.

## 3  Space charge forces

### 3.1  Direct space charge forces in free space

Let us consider a relativistic charge moving with constant velocity $\vec{v}$. It is well known that its electrostatic field is modified because of the relativistic Lorentz contraction along the direction of motion, as shown in Fig. 4. For an ultra-relativistic charge with $\gamma \to \infty$, the field lines are confined to a plane perpendicular to the direction of motion.

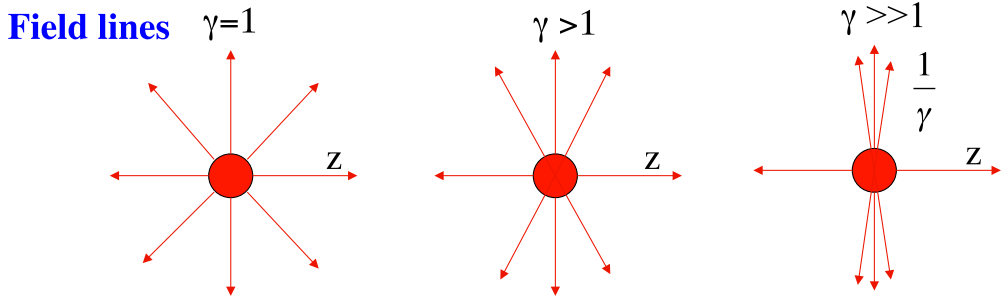

**Fig. 4:** Field lines for charges at different energies

If another charge is travelling on a parallel trajectory with respect to the first one along the $z$ axis, it is easy to see that the e.m. forces between them vanish. In fact, from the relativistic transforms of the electric and magnetic fields of a point charge, and using a cylindrical coordinate system $(r,\varphi,z)$ with the origin in the source charge, we have

$$E_r(z=0) = \frac{q}{4\pi\varepsilon_0}\frac{\gamma}{r^2},$$

$$B_\varphi(z=0) = \frac{q\beta}{4\pi\varepsilon_0 c}\frac{\gamma}{r^2}, \qquad (29)$$

$$E_z(r=0) = \frac{q}{4\pi\varepsilon_0}\frac{1}{\gamma^2 z^2}.$$

If the two charges travel along the $z$ axis at $r=0$ with different longitudinal positions, then the force is proportional to the longitudinal electric field $E_z$, and it vanishes as $1/\gamma^2$. On the other hand, if the charges have the same longitudinal position ($z=0$) and different transverse position, due to the combined effect of the defocusing electric and focusing magnetic fields we get

$$F_r = q\left(E_r - \beta c B_\varphi\right) = \frac{q\gamma}{4\pi\varepsilon_0 r^2}\left(1-\beta^2\right) = \frac{q}{4\pi\varepsilon_0 \gamma}\frac{1}{r^2}\,. \qquad (30)$$

In both cases, for $\gamma \to \infty$, a charge travelling close to another one on a parallel trajectory is not affected by e.m. forces.

Let us now consider the case of a uniform cylindrical charge distribution travelling with ultra-relativistic speed in free space. Under these assumptions, the electric field lines are perpendicular to the direction of motion, and the magnetic ones are circumferences, as shown in Fig. 5. The transverse electric and magnetic field intensities can be computed as in the static case, applying Gauss's and Ampère's laws:

$$\int_S \vec{E} \cdot \hat{n} \, dS = \frac{q}{\varepsilon_0}, \qquad \oint_l \vec{B} \cdot d\vec{l} = \mu_0 I . \tag{31}$$

We now suppose that the beam is a uniform cylinder of radius $a$ so that the longitudinal charge distribution (charge per unit of length) can be written as $\lambda(r) = \lambda_0 (r/a)^2$, and we want to compute the transverse space charge forces acting on a particle inside the beam.

Applying Eqs. (31) to a cylinder for Gauss's law and to a circumference for Ampère's law, we obtain

$$E_r(2\pi r)\Delta z = \frac{\lambda(r)\Delta z}{\varepsilon_0} \Rightarrow E_r = \frac{\lambda(r)}{2\pi\varepsilon_0 r} = \frac{\lambda_0}{2\pi\varepsilon_0} \frac{r}{a^2},$$

$$2\pi r B_\varphi = \mu_0 J \pi r^2 = \mu_0 \beta c \lambda(r) \Rightarrow B_\varphi = \frac{\lambda_0 \beta}{2\pi\varepsilon_0 c} \frac{r}{a^2}. \tag{32}$$

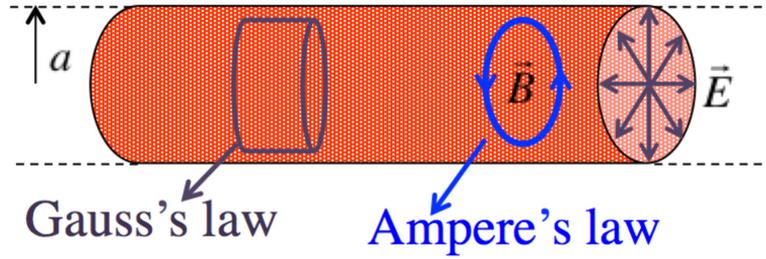

**Fig. 5:** Uniform cylindrical charge distribution with its electric and magnetic fields

We now observe that $B_\varphi = \frac{\beta}{c} E_r$, so that the e.m. transverse force acting on a charge inside the beam is given by

$$F_r(r) = e(E_r - \beta c B_\varphi) = e(1-\beta^2)E_r = \frac{e}{\gamma^2} \frac{\lambda_0}{2\pi\varepsilon_0} \frac{r}{a^2} . \tag{33}$$

We can therefore conclude that inside a uniform cylindrical charge density, travelling with ultra-relativistic speed, the transverse space charge forces vanish as $1/\gamma^2$ due to the cancellation of the electric and the magnetic forces.

## 3.2 Effects of conducting and magnetic screens

In an accelerator, the beams travel inside a vacuum pipe generally made of a metallic material (such as aluminium, copper, or stainless steel). This pipe also passes through the coils of the magnets (dipoles, quadrupoles, and sextupoles), and its cross-section may have a complicated shape, as in the case of special devices like RF cavities, kickers, diagnostics, and controls. However, most of the beam pipe has a cross-section with a simple shape: circular, elliptic, or quasi-rectangular. To obtain the space charge forces acting on a beam, let us consider a perfectly conducting, smooth beam pipe.

Before dealing with the problem, it is first necessary to review the basic features of e.m. fields close to metallic and magnetic materials. A discussion of the boundary conditions is given in Appendix A. We report only the more relevant conclusions here. The electric field of a point charge close to a conducting screen can be derived via the method of images, as shown in Fig. A.1. Concerning the magnetic field, a constant current close to a good conductor screen with $\mu_r \gg 1$, such as copper or aluminium, produces circular field lines that are not affected by the presence of the material itself. However, if the material is of ferromagnetic type, with $\mu_r \gg 1$, due to its magnetization the magnetic field lines are strongly affected inside and outside the material. In particular, a very high magnetic permeability makes the tangential field zero at the boundary so that the total magnetic field must be perpendicular to the surface, just as for the electric field lines close to a conductor (see Fig. A.2).

As discussed in Appendix A, the scenario changes when we deal with time-varying fields, for which it is necessary to compare the wall thickness and the skin depth (the region of penetration of the e.m. fields) in the conductor. If the fields penetrate and pass through the material, we are practically in the static boundary conditions case. Conversely, if the skin depth is very small, fields do not penetrate, and then the electric field lines are perpendicular to the wall, as in the static case, while the magnetic field lines are tangential to the surface. In this case, the magnetic field lines can be obtained by considering two currents flowing in opposite directions.

In the following we analyse the forces due to the presence of the screens in some simple cases.

### 3.3 Circular perfectly conducting pipe with beam at the centre and direct space charge forces

Due to the symmetry, the transverse fields produced by an ultra-relativistic charge inside a circular, perfectly conducting pipe are the same as those in free space. This implies that for a charge distribution with cylindrical symmetry, in the ultra-relativistic regime, the total force acting on a charge inside the beam is still given by Eq. (33). It is interesting to note that this result does not depend on the longitudinal distribution of the beam, so that, considering more generally a uniform radial distribution and a longitudinal linear density $\lambda(z)$, the force is given by

$$F_r(r,z) = \frac{e}{\gamma^2} \frac{\lambda(z)}{2\pi\varepsilon_0} \frac{r}{a^2}. \tag{34}$$

This force, like that in free space, has a dependence that goes as $1/\gamma^2$ due to the cancellation of the electric and magnetic forces, and it is linear with the transverse position $r$. If the transverse distribution is not constant, we can still apply Gauss's law to obtain the electric field and Ampere's law to obtain the magnetic field. For example, consider the following distribution:

$$\rho(r,z) = \frac{q_0}{\left(\sqrt{2\pi}\right)^3 \sigma_z \sigma_r^2} e^{-z^2/2\sigma_z^2} e^{-r^2/2\sigma_r^2}, \tag{35}$$

where $q_0$ is the bunch charge. Gauss's law, when applied to a cylinder with an infinitesimal height $dz$, as shown in Fig. 5, yields for the radial electric field (we suppose $\gamma \to \infty$ so that $E_z \approx 0$)

$$E_r(r,z) = \frac{1}{2\pi\varepsilon_0} \frac{q_0}{\sqrt{2\pi}\sigma_z \sigma_r^2 r} e^{-z^2/2\sigma_z^2} \int_0^r e^{-r'^2/2\sigma_r^2} r' dr' = \frac{1}{2\pi\varepsilon_0} \frac{q_0}{\sqrt{2\pi}\sigma_z} e^{-z^2/2\sigma_z^2} \left[\frac{1-e^{-r^2/2\sigma_r^2}}{r}\right]. \tag{36}$$

The magnetic field can be obtained in the same way as in Eq. (32) so that the total force on a charge inside the bunch is given by

$$F_r(r,z) = e(1-\beta^2)E_r = \frac{e}{2\pi\varepsilon_0\gamma^2}\frac{q_0}{\sqrt{2\pi}\sigma_z}e^{-z^2/2\sigma_z^2}\left[\frac{1-e^{-r^2/2\sigma r^2}}{r}\right]. \qquad (37)$$

It is important to observe that the self-induced forces given by Eqs. (34) and (37) are always defocusing in either the $x$ or the $y$ direction, as shown in Fig. 6. Note that the force given by Eq. (37) is not linear in the transverse position.

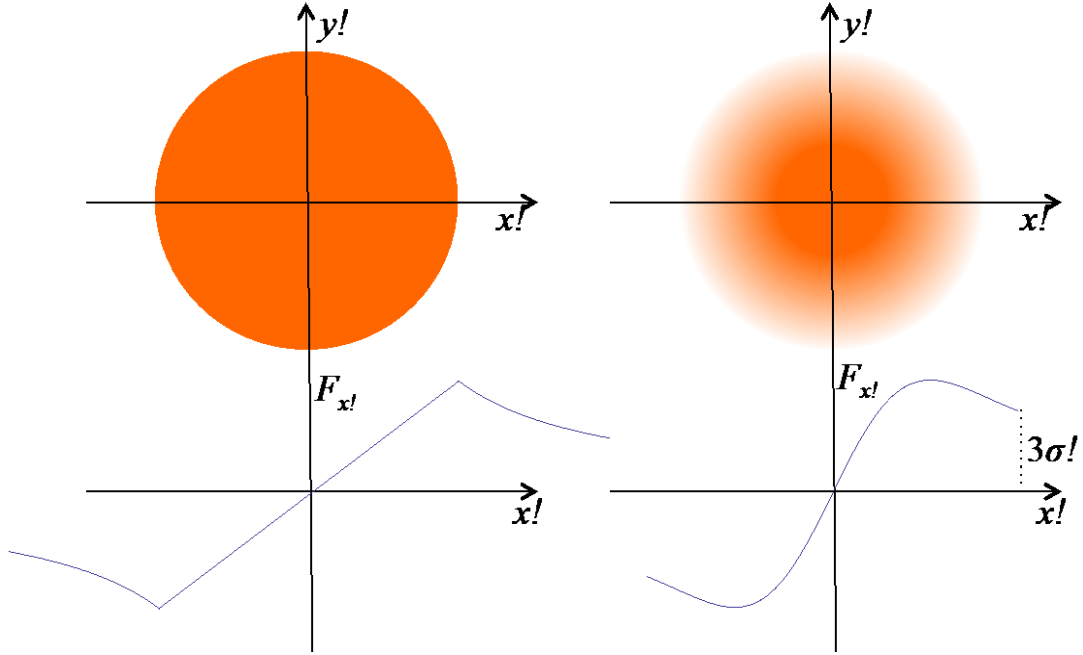

**Fig. 6:** Defocusing transverse self-induced forces produced by direct space charge in case of uniform (left) and Gaussian (right) distributions.

### 3.4 Parallel plates with beam at the centre and indirect space charge forces for d.c. currents

In some cases, and also with an elliptical beam pipe, the cross-section is such that we can consider the surfaces closer to the beam only, which can be approximated by two parallel plates. Let us suppose a charge distribution $\lambda(z)$ of radius $a$ between two conducting plates a distance $2h$ apart. In obtaining the static electric field, the two conducting plates can be removed by using the method of images and substituted by an infinite series of charges with alternating sign, $2h$ apart, as shown in Fig. 7.

We now want to evaluate the electric field, due to the image charges, at a position $y$ inside the bunch ($y < a$). The transverse field of the image charge distribution immediately above the real one can be written as

$$E_y^{1,\text{up,im}}(z,y) = \frac{\lambda(z)}{2\pi\varepsilon_0}\frac{1}{2h-y}, \qquad (38)$$

and the transverse electric field of the image charge distribution immediately below the real one is

$$E_y^{1,\text{down,im}}(z,y) = -\frac{\lambda(z)}{2\pi\varepsilon_0}\frac{1}{2h+y}. \qquad (39)$$

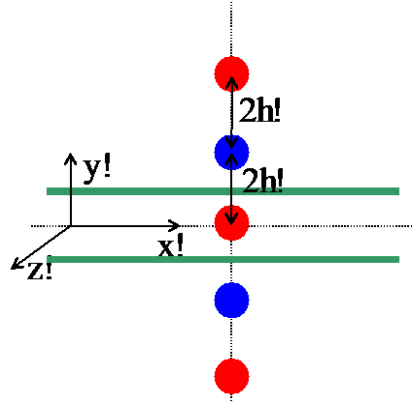

**Fig. 7:** A charge between two parallel plates and its image charges

If we sum the contribution of all the infinite image charge distributions, we get the total transverse electric field:

$$E_y^{im}(z,y) = \frac{\lambda(z)}{2\pi\varepsilon_0}\sum_{n=1}^{\infty}(-1)^n\left[\frac{1}{2nh+y}-\frac{1}{2nh-y}\right] = \frac{\lambda(z)}{2\pi\varepsilon_0}2y\sum_{n=1}^{\infty}(-1)^{n+1}\frac{1}{(2nh)^2-y^2}. \quad (40)$$

The sum on the right-hand side can be carried out. We consider, however, the simplified hypothesis that $h \gg a > y$, so we can ignore the $y$ term in the denominator of the sum to get

$$E_y^{im}(z,y) \cong \frac{\lambda(z)}{2\pi\varepsilon_0}\frac{y}{2h^2}\sum_{n=1}^{\infty}\frac{(-1)^{n+1}}{n^2} = \frac{\lambda(z)}{4\pi\varepsilon_0 h^2}\frac{\pi^2}{12}y. \quad (41)$$

For d.c. or slowly varying currents, we have seen that the boundary conditions imposed by the conducting plates do not affect the magnetic field, which remains circular with no image currents. As a consequence there is no cancellation effect of the electric and magnetic forces for the fields produced by the images, as we obtained for the real charges (direct forces), and the indirect force acting on a charge inside the beam is simply the electric field given by Eq. (41) multiplied by the particle charge.

From the divergence equation, we can also derive the other transverse component of the electric field along $x$:

$$\frac{\partial}{\partial x}E_x^{im} = -\frac{\partial}{\partial y}E_y^{im} \Rightarrow E_x^{im}(z,x) = \frac{-\lambda(z)}{4\pi\varepsilon_0 h^2}\frac{\pi^2}{12}x. \quad (42)$$

From the above fields, the total forces acting on a charge inside the bunch moving between two parallel plates, including the direct space charge force given by Eq. (34), are given by

$$F_x(z,x) = \frac{e\lambda(z)}{\pi\varepsilon_0}\left(\frac{1}{2a^2\gamma^2} - \frac{\pi^2}{48h^2}\right)x \quad (43)$$

$$F_y(z,y) = \frac{e\lambda(z)}{\pi\varepsilon_0}\left(\frac{1}{2a^2\gamma^2} + \frac{\pi^2}{48h^2}\right)y. \quad (44)$$

Therefore, for $\gamma \ll 1$, and for d.c. or slowly varying currents, the cancellation effect applies only to the direct space charge forces. There is no cancellation of the electric and magnetic forces due to the image charges.

## 3.5 Parallel plates with beam at the centre and indirect space charge forces for a.c. currents

We have seen that close to a conductor the e.m. fields have different behaviours, depending on the skin depth $\delta_w$ of the material (Appendix A). Usually, the frequency spectrum of a beam is quite rich in harmonics, especially for bunched beams. It is then convenient to decompose the current into a d.c. component, $\bar{I}$, for which $\delta_w \gg \Delta w$, where $\Delta w$ is the width of the beam pipe, and an a.c. component, $\hat{I}$, for which $\delta_w \ll \Delta w$. While the d.c. component of the magnetic field does not perceive the presence of the material, so that we can apply Eqs. (43) and (44), its a.c. component produces a magnetic field tangential to the wall, which can be obtained by using an infinite sum of image currents with alternating directions. In this case, to obtain the total magnetic field of the image currents, we can follow the same procedure we used for the electric fields given by Eqs. (41) and (42), and by considering the relation between the a.c. current and its charge distribution: $\hat{I} = \beta c \hat{\lambda}$, thus obtaining a magnetic field due to image currents of the following kind:

$$\hat{B}_x(z,y) = -\frac{\beta}{c}\hat{E}_y(z,y) = -\frac{\beta^2 \hat{\lambda}(z)}{\pi \varepsilon_0} \frac{\pi^2}{48h^2} y. \tag{45}$$

We can see from Eq. (45) that in this case the attractive magnetic force tends to compensate for the repulsive electric one, which is still given by the second term in the parentheses of Eq. (44), so that we have a total force due to image charges and currents given by

$$\hat{F}_y(z,y) = \frac{e\hat{\lambda}(z)}{\pi \varepsilon_0 \gamma^2} \frac{\pi^2}{48h^2} y. \tag{46}$$

Combining Eq. (46) with the direct space charge force, we get, for the a.c. component,

$$\hat{F}_y(z,y) = \frac{e\hat{\lambda}(z)}{2\pi \varepsilon_0 \gamma^2}\left(\frac{1}{a^2} + \frac{\pi^2}{24h^2}\right)y, \tag{47}$$

and analogously, along the $x$ direction,

$$\hat{F}_x(z,x) = \frac{e\hat{\lambda}(z)}{2\pi \varepsilon_0 \gamma^2}\left(\frac{1}{a^2} - \frac{\pi^2}{24h^2}\right)x. \tag{48}$$

## 3.6 Parallel plates with beam at the centre and indirect space charge forces for d.c. currents in the presence of ferromagnetic materials

As a final example, we consider the case where there is a dipole magnet outside the metallic pipe. The magnetic field produced by the d.c. currents, $\beta c \bar{\lambda}(z)$, does not see the conducting pipe, although it is strongly affected by the ferromagnetic material. In fact, as seen in Appendix A, the magnetic field lines must be orthogonal to the pole surface. We have also seen that the total magnetic field can be obtained by removing the screen and considering image currents flowing in the same direction.

Proceeding analogously to Section 3.4, where $g$ is the gap in the dipole magnet, we obtain

$$B_x^{im}(z,y) = \frac{\mu_0 \beta c \bar{\lambda}(z)}{2\pi} \sum_{n=1}^{\infty}\left[\frac{1}{2ng-y} - \frac{1}{2ng+y}\right]. \tag{49}$$

Note that in this case we do not have the $(-1)^n$ term. By using the same approximation as before ($h \gg a > y$), we obtain a magnetic field due to the image currents equal to

$$B_x^{im}(z,y) \cong \frac{\mu_0 \beta c \overline{\lambda}(z) y}{4\pi g^2} \sum_{n=1}^{\infty} \frac{1}{n^2} = \frac{\mu_0 \beta c \overline{\lambda}(z) \pi^2 y}{24\pi g^2}, \qquad (50)$$

and the corresponding force is given by

$$F_y^{im}(z,y) = \frac{\beta^2 \overline{\lambda}(z) \pi^2}{24\pi \varepsilon_0 g^2} y. \qquad (51)$$

To obtain the magnetic field acting on a particle displaced along the $x$ direction, we can use the relation $\vec{\nabla} \times \vec{B} = 0$, which for the $z$ direction gives

$$\frac{\partial B_y}{\partial x} = \frac{\partial B_x}{\partial y}, \qquad (52)$$

so that

$$B_y^{im}(z,x) = \frac{\mu_0 \beta c \overline{\lambda}(z) \pi^2}{24\pi g^2} x \qquad (53)$$

and

$$F_x^{im}(z,x) = -\frac{\beta^2 \overline{\lambda}(z) \pi^2}{24\pi \varepsilon_0 g^2} x. \qquad (54)$$

### 3.7  Parallel plates with beam at the centre: general expression of the force

Taking into account all the boundary conditions and either d.c. and a.c. currents, we can summarize what we have obtained in the preceding sections for the parallel plates, and write the general expression of the force as follows:

$$F_u = \frac{e}{2\pi \varepsilon_0} \left[ \frac{1}{\gamma^2} \left( \frac{1}{a^2} \mp \frac{\pi^2}{24 h^2} \right) \lambda \mp \beta^2 \left( \frac{\pi^2}{24 h^2} + \frac{\pi^2}{12 g^2} \right) \overline{\lambda} \right] u, \qquad (55)$$

where $\lambda$ is the total current divided by $\beta c$, and $\overline{\lambda}(z)$ is its d.c. term. We take the (+) sign if $u = y$, and the (−) sign if $u = x$.

One often sees Eq. (55) written as follows:

$$F_u = \frac{e}{\pi \varepsilon_0} \left[ \frac{1}{\gamma^2} \left( \frac{\varepsilon_0}{a^2} \mp \frac{\varepsilon_1}{h^2} \right) \lambda \mp \beta^2 \left( \frac{\varepsilon_1}{h^2} + \frac{\varepsilon_2}{g^2} \right) \overline{\lambda} \right] u, \qquad (56)$$

where the Laslett form factors [10], $\varepsilon_0$, $\varepsilon_1$, and $\varepsilon_2$, can be obtained for several beam pipe geometries. For example, for parallel plates, on comparing Eq. (55) with Eq. (56) we get $\varepsilon_0 = 1/2$, $\varepsilon_1 = \pi^2/48$, and $\varepsilon_2 = \pi^2/24$ (see, for example, Ref. [11]). It is interesting to note that these forces are linear anyway in the transverse displacements $x$ and $y$.

### 3.8  Longitudinal direct space charge force

Up to now we have obtained the direct and indirect transverse forces produced by space charge distributions. The longitudinal electric field, which is responsible for the longitudinal forces, can be derived starting from knowledge of the transverse fields, as shown in Appendix B. The transverse electric field inside the beam ($r \leq a$) can be expressed by the first of Eqs. (32) for a uniform transverse distribution, which, however, can be generalized by considering a non-uniform longitudinal distribution $\lambda(z)$, and outside the beam ($r \geq a$) it is equal to

$$E_r(r \geq a) = \frac{\lambda(z)}{2\pi\varepsilon_0 r}. \tag{57}$$

As a consequence, the final equation of Appendix B becomes

$$E_z(r,z) = -\frac{1}{2\pi\varepsilon_0 \gamma^2}\left[\int_r^a \frac{r'}{a^2}\,\mathrm{d}r' + \int_a^b \frac{1}{r'}\,\mathrm{d}r'\right]\frac{\partial \lambda(z)}{\partial z}, \tag{58}$$

giving a longitudinal force of the following kind:

$$F_z(r,z) = \frac{-e}{4\pi\varepsilon_0 \gamma^2}\left(1 - \frac{r^2}{a^2} + 2\ln\frac{b}{a}\right)\frac{\partial \lambda(z)}{\partial z}. \tag{59}$$

Therefore, the longitudinal force acting on a charge is positive (negative) in the region with negative (positive) density slope.

## 4    Coherent and incoherent tune shifts

### 4.1    Coherent and incoherent effects

We are now ready to study the effects of the space charge forces on the beam dynamics. When the beam is located at the centre of symmetry of the pipe, the e.m. forces due to direct space charge and image currents cannot affect the motion of the centre of mass (coherent motion), but they can change the trajectory of individual charges inside the beam (incoherent effects). These forces may have a complicated dependence on the charge position. A simple analysis is performed, considering only the linear expansion of the self-induced forces around the equilibrium trajectory, as the forces given by Eq. (34) or Eq. (55).

Referring to the equations of motion, Eqs. (22), (23), and (28), we now focus our attention on the self-induced forces and expand them around the ideal orbit analogously to that for the external forces. A constant term in the expansion of $F^{\text{self}}$ changes the equilibrium orbit in the transverse plane, and the synchronous phase in the longitudinal plane, while the linear term, which is proportional to the displacement, changes the focusing strength and therefore induces a shift of the betatron and synchrotron frequencies.

This can happen either in the motion of individual particles inside the beam (incoherent motion) or in the transverse oscillations of the whole beam (coherent motion) around the closed orbit when the beam is off-centre with respect to the beam pipe.

### 4.2    Transverse incoherent effects

Let us consider only the linear term of the transverse self-induced forces, i.e.

$$F_x^{\text{self}}(x,s) \cong \left(\frac{\partial F_x^{\text{self}}}{\partial x}\right)_{x=0} x. \tag{60}$$

For the case of a uniform transverse beam distribution, either in a circular pipe or between parallel plates, the force given by Eq. (60) is not an approximation, as shown by Eqs. (34) and (55). For other kinds of distributions, such as the one given by Eq. (35), we can always suppose that the transverse displacement $r$ of a charge is much smaller than the transverse bunch dimension $\sigma_r$, so that the term inside the square brackets in Eq. (37) can be expanded to first order in $r$, giving

$r/(2\sigma_r)$. In any case, we end up with a self-induced force linearly dependent on the particle displacement. As a consequence, by considering, for example, the motion along the $x$ direction, Eq. (22) becomes

$$x''(s) + \left(\frac{Q_x}{\rho_x}\right)^2 x(s) = \frac{1}{m_0 v_z^2 \gamma}\left(\frac{\partial F_x^{\text{self}}}{\partial x}\right)_{x=0} x(s). \tag{61}$$

The linear additional term on the right-hand side produces a shift of the betatron tune $Q_x$. Indeed, we can write

$$x''(s) + \left[\left(\frac{Q_x}{\rho_x}\right)^2 - \frac{1}{\beta^2 E_0}\left(\frac{\partial F_x^{\text{self}}}{\partial x}\right)_{x=0}\right] x(s) = 0, \tag{62}$$

where the term $m_0 v_z^2 \gamma$ has been substituted with $\beta^2 E_0$. We now recognize in the brackets a term proportional to the square of the new betatron tune, shifted with respect to the initial $Q_x$ by the self-induced forces. This term can be written as $(Q_x + \Delta Q_x)^2 / \rho_x^2$. For small perturbations, the shift $\Delta Q_x$ can then be obtained from

$$\left[\left(\frac{Q_x}{\rho_x}\right)^2 - \frac{1}{\beta^2 E_0}\left(\frac{\partial F_x^{\text{self}}}{\partial x}\right)_{x=0}\right] = \frac{(Q_x + \Delta Q_x)^2}{\rho_x^2} \cong \frac{Q_x^2 + 2Q_x \Delta Q_x}{\rho_x^2}, \tag{63}$$

thus giving

$$\Delta Q_x = -\frac{\rho_x^2}{2\beta^2 E_0 Q_x}\left(\frac{\partial F_x^{\text{self}}}{\partial x}\right). \tag{64}$$

A similar expression is found in the $y$ direction. The betatron tune shift is negative because the space charge forces are defocusing on both planes. Note that the tune shift is in general a function of $z$, due to the dependence of the self-induced force on $\lambda(z)$. The consequence is a tune spread inside the beam. This conclusion is generally also true for more realistic non-uniform transverse beam distributions, which are characterized by a tune shift dependent also on the betatron oscillation amplitude. In these cases, the effect is called tune spread instead of tune shift.

As an example of an application of Eq. (64), let us find the incoherent shift of the betatron tune for a uniform electron beam of charge $eN_p$, radius $a$, and length $l_0$, inside a circular pipe. The self-induced force is given by Eq. (34), with $\lambda(z) = eN_p/l_0$, so that

$$\Delta Q_x = -\frac{\rho_x^2 e^2 N_p}{4\pi\varepsilon_0 a^2 l_0 \beta^2 \gamma^2 E_0 Q_x}; \tag{65}$$

or, expressed in terms of the classical radius of electron $r_0$,

$$\Delta Q_x = -\frac{r_0 \rho_x^2 N_p}{a^2 l_0 \beta^2 \gamma^3 Q_x}. \tag{66}$$

In the general case of non-uniform focusing along the accelerator, as given by Eq. (16), the linear effect of the self-induced forces can be treated as a quadrupole error $\Delta K_u$ (see, for example, Ref. [11]) distributed along the accelerator, with $u$ representing either the $x$ or the $y$ axis, thus giving a betatron tune shift of

$$\Delta Q_u = \frac{1}{4\pi} \oint \beta_u(s) \Delta K_u(s) ds = \frac{-1}{4\pi\beta^2 E_0} \oint \beta_u(s) \left( \frac{\partial F_u^{\text{self}}}{\partial u} \right) ds. \tag{67}$$

For example, by considering the previous case of a uniform electron beam inside a circular pipe, but with non-uniform focusing, Eq. (65) will be replaced by

$$\Delta Q_x = -\frac{r_0 N_p}{2\pi\beta^2\gamma^3 l_0} \oint \frac{\beta_x(s)}{a^2(s)} ds = -\frac{r_0 N_p}{2\pi\beta^2\gamma^3 l_0} \frac{2\pi\rho_x}{\varepsilon_x}, \tag{68}$$

which has been obtained by observing that the quantity $a^2(s)/\beta(s)$ is the beam emittance $\varepsilon_x$, which is constant along the machine.

### 4.3 Transverse coherent effects

Let us suppose that a beam is displaced from the pipe axis, due for example to coherent betatron oscillations. Due to induction, there will be a higher concentration of charges of opposite sign on the pipe surface closer to beam, which attracts the beam itself more intensely than the induced charges on the opposite side. As a consequence, its centre of mass will experience a defocusing force.

As example, let us consider an electron beam with uniform longitudinal charge distribution $\lambda_0 = eN_p/l_0$ inside a conducting cylindrical pipe of radius $b$, displaced by $x$ from its axis, as shown in Fig. 8. Since the electric field lines of the charge distribution must be perpendicular to the pipe, by imposing that the conductor surface is equipotential we can remove and substitute the pipe with a charge distribution $-\lambda_0$ at a distance on the axis equal to $d = b^2/x$ in the same direction as the displacement.

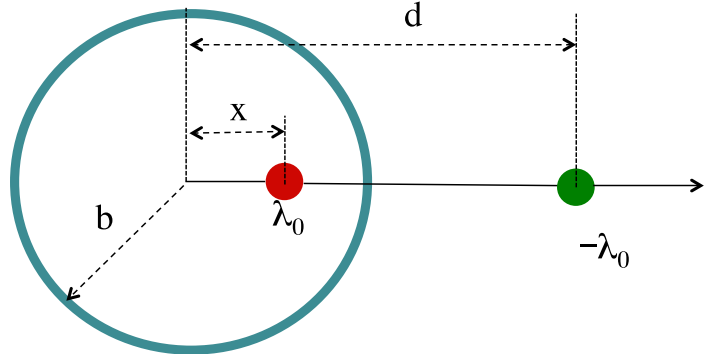

**Fig. 8:** Charge distribution inside a cylindrical pipe and its image charge distribution

The image charge distribution attracts the whole beam, thus producing a coherent defocusing effect. The electric field of the image charge acting on the centre of mass of the beam is given by

$$E_x^{\text{im}}(x) \cong \frac{\lambda_0}{2\pi\varepsilon_0} \frac{1}{d-x}. \tag{69}$$

If we linearize the electric field by considering small displacements of the beam, such that $x \ll d$, then $1/(d-x) \cong 1/d = x/b^2$, and we obtain

$$E_x^{\text{im}}(x) \cong \frac{\lambda_0}{2\pi\varepsilon_0} \frac{x}{b^2}, \tag{70}$$

which gives a linear coherent force

$$F_x^{\text{self}}(x) \cong \frac{e\lambda_0}{2\pi\varepsilon_0} \frac{x}{b^2}. \tag{71}$$

This force produces a coherent betatron tune shift that can still be evaluated by using Eq. (64), giving

$$\Delta Q_x = -\frac{\rho_x^2 e\lambda_0}{4\pi\varepsilon_0 \beta^2 E_0 b^2 Q_x} = -\frac{r_0 \rho_x^2 N_p}{b^2 l_0 \beta^2 \gamma Q_x}. \tag{72}$$

This coherent betatron tune shift, unlike the incoherent one given by Eq. (66), does not depend on the beam size but rather on the pipe radius, and it is inversely proportional to the beam energy.

### 4.4 Longitudinal incoherent effects

The effects of longitudinal space charge forces on the beam dynamics can be obtained by using Eq. (28). We expand the self-induced force and, as already discussed in Section 4.1, we observe that the constant term in the expansion leads to a shift of the synchronous phase, while the linear term, which is proportional to the displacement, changes the focusing strength and therefore induces a shift of the synchrotron tune.

Let us analyse the motion of the beam around the new equilibrium phase and consider the linear term of the longitudinal self-induced force:

$$F_z^{\text{self}}(r,z) \cong \left(\frac{\partial F_z^{\text{self}}}{\partial z}\right)_{z=0} z. \tag{73}$$

The equation of motion, Eq. (28), becomes

$$z'' + \left(\frac{Q_z}{\rho_x}\right)^2 z = \frac{\eta}{\beta^2 E_0} \left(\frac{\partial F_z^{\text{self}}}{\partial z}\right)_{z=0} z. \tag{74}$$

As in the transverse case, consider the same approximation leading to Eq. (63); in this case we obtain

$$\Delta Q_z = \frac{-\eta \rho_x^2}{2\beta^2 E_0 Q_z} \left(\frac{\partial F_z^{\text{self}}}{\partial z}\right). \tag{75}$$

Unlike for the transverse betatron tune shifts, the synchrotron tune shift can be either positive or negative, and can change with the position of the charge inside the beam.

For example, consider a transverse uniform beam of radius $a$ in a cylindrical pipe, having a parabolic longitudinal distribution of the type

$$\lambda(z) = \frac{3eN_p}{2l_0}\left[1 - \left(\frac{2z}{l_0}\right)^2\right]. \tag{76}$$

The incoherent tune shift can be obtained by combining Eqs. (59), (75), and (76), yielding

$$\Delta Q_z = -6\frac{\eta \rho_x^2 r_0 N_p}{\beta^2 \gamma^3 Q_z}\left(1 - \frac{r^2}{a^2} + 2\ln\frac{b}{a}\right)\left(\frac{1}{l_0^3}\right), \tag{77}$$

which depends on the transverse position $r$ of the charge inside the beam. If, instead of a parabolic bunch distribution, we had a Gaussian one, then the synchrotron tune shift would have a dependence also on the longitudinal position $z$.

## 5    Consequences of the space charge tune shifts

In circular accelerators the values of the betatron tunes should not be close to rational numbers to avoid the crossing of linear and non-linear resonances where the beam becomes unstable. The tune spread induced by the space charge force can make it difficult to satisfy this basic requirement. Typically, to avoid major resonances the stability requires [9, 12]

$$\left|\Delta Q_u\right| < 0.5 .$$

If the tune spread exceeds this limit, it is possible to reduce the effect of space charge tune spread by increasing the injection energy.

It is worth noting that the incoherent tune spread produces also a beneficial effect, called Landau damping, which can cure the coherent instabilities, provided that the coherent tune remains inside the incoherent spread.

## 6    Direct space charge effects in a linac

In a linac or beam transport line, direct space charge effects can lead to significant longitudinal–transverse correlations of the bunch parameters, which may produce mismatch with the focusing and accelerating devices, thus contributing to emittance growth (and energy spread). Matching conditions suitable for preserving the beam quality can be derived from a simple model, as will be shown in the following. A more detailed discussion can be found in the many classical textbooks on this subject, such as the ones listed in Refs. [3] and [13].

Let us consider a *bunched* beam with initially uniform charge distribution in a cylinder of radius $R$ and length $l_0$, carrying a current $I$ and moving with longitudinal velocity $v = \beta c$. The linear components of the longitudinal and transverse space charge fields are given by [14]

$$E_z(\zeta) = \frac{IL}{2\pi\varepsilon_0 R^2 \beta c} h(\zeta), \tag{78}$$

$$E_r(r,\zeta) = \frac{Ir}{2\pi\varepsilon_0 R^2 \beta c} g(\zeta), \tag{79}$$

where the field form factors are described by the following functions:

$$h(\zeta) = \sqrt{A + (1-\zeta)^2} - \sqrt{A + \zeta^2} + (2\zeta - 1), \tag{80}$$

$$g(\zeta) = \frac{(1-\zeta)}{2\sqrt{A^2 + (1-\zeta)^2}} + \frac{\zeta}{2\sqrt{A^2 + \zeta^2}}; \tag{81}$$

$\zeta = z/L$ is the normalized longitudinal coordinate along the bunch, and $A = R/\gamma L$ is the beam aspect ratio. As $\gamma$ increases, $g(\zeta) \to 1$ and $h(\zeta) \to 0$. Thus, direct space charge fields mainly affect transverse beam dynamics.

The transverse beam dynamics of a beam characterized by an rms envelope $\sigma = \sqrt{\langle x^2 \rangle}$ and transverse normalized thermal rms emittance at the source [3],

$$\varepsilon_{n,th}^2 = \frac{\langle x^2 \rangle \langle p_x^2 \rangle}{(m_0 c)^2} = \frac{\gamma k_B T \sigma_0^2}{m_0 c^2}, \tag{82}$$

can be conveniently described, under the paraxial ray approximation, i.e. $p_x \ll p_z$, by the rms envelope equation for an axisymmetric beam [3, 15, 16]:

$$\sigma'' + \frac{\gamma'}{\gamma}\sigma' + k_{ext}^2 \sigma = \frac{K_{sc}}{\gamma^3 \sigma} + \frac{\varepsilon_{n,th}^2}{\gamma^2 \sigma^3}. \tag{83}$$

The first term is the change in the envelope slope; the second term drives the envelope oscillation damping due to acceleration; the third term accounts for linear external focusing forces; the fourth term represents the defocusing space charge effects; and the fifth term describes the internal pressure due to the emittance. Note that $K_{sc} = \hat{I}/2I_A$ is the *beam perveance*, where $\hat{I}$ is the peak current and $I_A$ is the Alfvén current (~17 kA), and $\gamma' = eE_{acc}/mc^2$, $E_{acc}$ being the accelerating field.

From the envelope equation (83) we can identify two regimes of beam propagation: *space charge dominated* and *emittance dominated*. A beam is space charge dominated as long as the space charge collective forces are largely dominant over the emittance pressure. A measure of the relative importance of space charge effects versus emittance pressure is given by the *laminarity parameter*, defined as the ratio between the space charge term and the emittance term:

$$\rho = \frac{\hat{I}}{2I_A \gamma} \frac{\sigma^2}{\varepsilon_n^2}. \tag{84}$$

When $\rho$ greatly exceeds unity, the beam behaves like a laminar flow (all beam particles move on trajectories that do not cross), and transport and acceleration require a careful tuning of focusing and accelerating elements in order to retain the laminarity. Correlated emittance growth is typical in this regime, which can conveniently be made reversible if proper beam matching conditions are fulfilled, as discussed in the following. When $\rho \approx 1$ the beam is emittance dominated (this is also referred to as the thermal regime, corresponding to a Debye length large compared to the bunch envelope) and the space charge effects can be neglected. The transition to the thermal regime occurs when $\rho \approx 1$, corresponding to the transition energy

$$\gamma_{tr} = \frac{\hat{I}}{2I_A} \frac{\sigma^2}{\varepsilon_n^2}. \tag{85}$$

For example, a beam with $\hat{I}$ =100 A, $\varepsilon_n$ =1 $\mu$m, and $\sigma$ = 300 $\mu$m is leaving the space-charge-dominated regime and is entering the thermal regime at a transition energy of 131 MeV. From this example, one may conclude that space-charge-dominated regime is typical of low-energy beams. Actually, for applications like linac-driven free-electron lasers, high-density beams with peak currents in excess of kiloamps are required. Even if the bunch energy has reached values higher than $\gamma_{tr}$, space charge effects may recur if bunch compressors are actively increasing $\hat{I}$, so that a new transition energy with higher $\hat{I}$ has to be considered.

When longitudinal correlations within the bunch are important, such as the one induced by the space charge effects, the beam envelope evolution is generally dependent also on the longitudinal bunch coordinate $\zeta$. In this case, the bunch should be considered as an ensemble of $N$ longitudinal slices of envelope $\sigma_s(z,\zeta)$, whose evolution can be computed considering $N$ slice envelope equations equivalent to Eq. (83) provided that the bunch parameters refer to each single slice: $\gamma_s, \gamma'_s, k_{sc,s} = k_{sc} g(\zeta)$. Correlations within the bunch may cause emittance oscillations that can be evaluated, once a solution of the slice envelope equation is known, by using the following correlated emittance definition:

$$\varepsilon_{n,rms,cor} = \langle \gamma \rangle \sqrt{\left(\langle \sigma_s^2 \rangle \langle \sigma'^2_s \rangle - \langle \sigma_s \sigma'_s \rangle\right)}, \tag{86}$$

where the average is performed over the entire slice ensemble. In the simplest case of only two slices, the previous definition reduces to

$$\varepsilon_{n,rms,cor} = \langle \gamma \rangle |\sigma_1 \sigma'_2 - \sigma_2 \sigma'_1|, \tag{87}$$

which is a simple and useful formula for the estimation of the emittance scaling [17].

The total normalized rms emittance is then given by the superposition of the correlated and uncorrelated terms:

$$\varepsilon_{n,rms} = \sqrt{\left(\varepsilon_{n,th}^2 + \varepsilon_{n,rms,cor}^2\right)}. \tag{88}$$

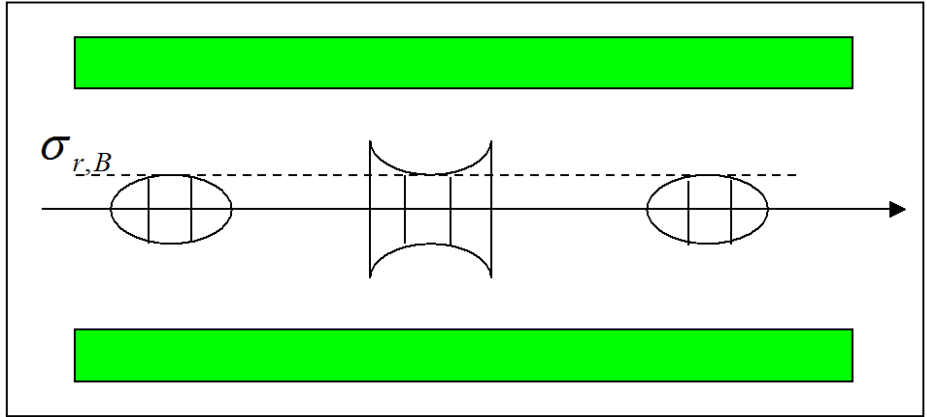

**Fig. 9:** Schematic representation of a nearly matched beam in a long solenoid. The dashed line represents the reference slice envelope fully matched to the Brillouin flow condition. The other slice envelopes are oscillating around the equilibrium solution.

An interesting example [6] to consider here, showing the consequences of a non-perfect beam matching, is the propagation of a beam in the space-charge-dominated regime nearly matched to an external focusing channel produced by a long solenoid, giving

$$k_{ext}^2 = k_{sol}^2 = \left(\frac{qB}{2\gamma m_0 c}\right)^2,$$

as illustrated in Fig. 9. To simplify our analysis we can neglect acceleration, as in the case of a simple beam transport line. The envelope equation for each slice, indicated by $\sigma_s$, reduces to

$$\sigma''_s + k_{ext}^2 \sigma_s = \frac{k_{sc,s}}{\gamma^3 \sigma_s}. \tag{89}$$

A stationary solution, called *Brillouin flow*, is given by

$$\sigma_{s,B} = \frac{1}{k_{ext}^2}\sqrt{\frac{\hat{I}g(\zeta)}{2\gamma^3 I_A}}, \qquad (90)$$

where the local dependence of the current $\hat{I}_s = \hat{I}g(\zeta)$ within the bunch has been explicitly indicated. This solution represents the matching conditions for which the external focusing completely balances the internal space charge force. Unfortunately, since $k_{ext}$ has a slice-independent constant value, the Brillouin matching condition cannot be achieved at the same time for all the bunch slices. Assuming there is a reference slice perfectly matched with an envelope $\sigma_{r,B}$, the matching condition for the other slices can be written as follows:

$$\sigma_{s,B} = \sigma_{r,B} + \frac{\sigma_{r,B}}{2}\left(\frac{\delta I_s}{\hat{I}}\right), \qquad (91)$$

with respect to the reference slice. Considering a small perturbation $\delta_s$ from the equilibrium in the form

$$\sigma_s = \sigma_{s,B} + \delta_s \qquad (92)$$

and substituting into Eq. (89), we obtain a linearized equation for the slice offset:

$$\delta_s'' + 2k_{ext}^2 \delta_s = 0, \qquad (93)$$

which has a solution given by

$$\delta_s = \delta_0 \cos\left(\sqrt{2}k_{ext}z\right), \qquad (94)$$

where $\delta_0 = \sigma_{s,0} - \sigma_{s,B}$ is the amplitude of the initial slice mismatch that we assume for convenience to be the same for all slices. Inserting Eq. (94) into Eq. (92), we get the perturbed solution:

$$\sigma_s = \sigma_{s,B} + \delta_0 \cos\left(\sqrt{2}k_{ext}z\right). \qquad (95)$$

Equation (95) shows that slice envelopes oscillate together around the equilibrium solution with the same frequency for all slices ($\sqrt{2}k_{ext}$, often called the plasma frequency) which is dependent only on the external focusing forces. This solution represents a collective behaviour of the bunch, similar to that demonstrated by electrons subject to the restoring force of ions in a plasma. Using the two-slice model and Eq. (95), the emittance evolution Eq. (87) results:

$$\varepsilon_{n,rms,cor} = \frac{1}{4}\langle\gamma\rangle k_{ext}\sigma_{r,B}\left|\frac{\Delta I}{\hat{I}}\delta_0 \sin\left(\sqrt{2}k_{ext}z\right)\right|, \qquad (96)$$

where $\Delta I = \hat{I}_1 - \hat{I}_2$. Note that in this simple case envelope oscillations of the mismatched slices induce correlated emittance oscillations that periodically go back to zero, showing the reversible nature of the correlated emittance growth. It is, in fact, the coupling between the transverse and longitudinal motion induced by the space charge fields that allows reversibility. With a proper tuning of either the transport line length or the focusing field, one can compensate for the transverse emittance growth. Similar arguments can be considered when including acceleration, despite the fact that the analytical treatment results are more complex [6]. In this case, for a beam well matched to the accelerating structure, the plasma frequency $\sqrt{2}K_{ext} \to 0$ with increasing $\gamma$. Before the transition energy is achieved, the emittance performs damped oscillations. Careful tuning of the external fields can minimize the value of the emittance at the beam extraction.

# Appendix A: Boundary relations for conductors

## A.1 Static electric and magnetic fields

When we have two materials with different relative permittivities $\varepsilon_{r1}$ and $\varepsilon_{r2}$, the tangential electric field and the normal electric displacement are preserved in the transition from one material to another, so that we have the following boundary relations:

$$E_{t1} = E_{t2},$$
$$\varepsilon_{r1} E_{n1} = \varepsilon_{r2} E_{n2}.$$

If one of the two materials is a conductor with a finite conductivity, the electric field vanishes inside it, and the walls are equipotential surfaces. This implies that the electric field lines are orthogonal to the conductor surface, independent of the dielectric and magnetic properties of the material. The only condition is that there is a finite conductivity.

In the case of a charge close to a conductor, to obtain the electric field we need to include the effects of the induced charges on the conducting surfaces, and we must know how they are distributed. Generally this task is not easy, but if we have an infinite conducting screen the problem can be easily solved by making use of the method of images: we can remove the screen and put at a symmetric location a charge with opposite sign, as shown in Fig. A.1.

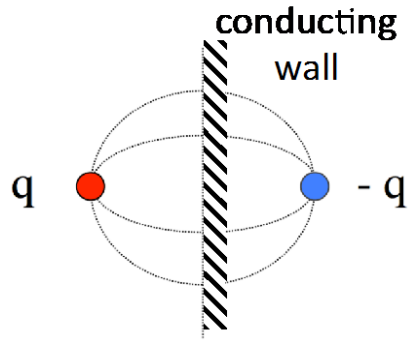

**Fig. A.1:** Method of images

The total electric field is the sum of the direct and the image fields:

$$\vec{E}^{tot} = \vec{E}^{direct} + \vec{E}^{images}.$$

For the static magnetic field between two materials with different permeabilities, the following boundary relations hold:

$$H_{t1} = H_{t2},$$
$$\mu_{r1} H_{n1} = \mu_{r2} H_{n2}.$$

Thus, static magnetic fields do not perceive the presence of the conductor if it has a magnetic permeability $\mu_r \approx 1$, as for copper or aluminium, and the field lines expand as in free space. However, a beam pipe in a real machine goes through many magnetic components (such as dipoles and quadrupoles) made of ferromagnetic materials with high permeabilities (of the order $10^3$–$10^5$). For these materials, due to the boundary conditions, the magnetic field lines are practically orthogonal to the surface. Similar to electric field lines for a conductor, the total magnetic field can be derived by using the image method: we remove the magnetic wall and put a symmetric current with the same sign, as shown in Fig. A.2.

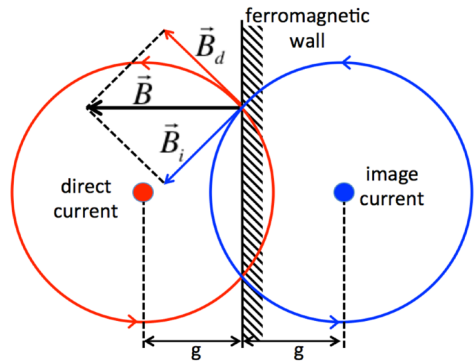

**Fig. A.2:** Method of image current

## A.2  Time-varying fields

Static electric fields vanish inside a conductor for any finite conductivity, whereas static magnetic fields pass through unless the conductor has a high permeability. This is not true for time-varying fields, which can penetrate the material in a region called the skin depth, $\delta_w$. To present the skin depth as a function of the material properties, we write the following Maxwell equations inside the conducting material together with the constitutive relations

$$\begin{cases} \nabla \times \vec{E} = -\dfrac{\partial \vec{B}}{\partial t} \\ \nabla \times \vec{H} = \vec{J} + \dfrac{\partial \vec{D}}{\partial t} \end{cases} \qquad \begin{cases} \vec{B} = \mu \vec{H} \\ \vec{D} = \varepsilon \vec{E} \\ \vec{J} = \sigma \vec{E} \end{cases}.$$

Let us consider a plane wave linearly polarized with the electric field in the $y$ direction propagating in the material along the $x$ direction, as shown in Fig. A.3.

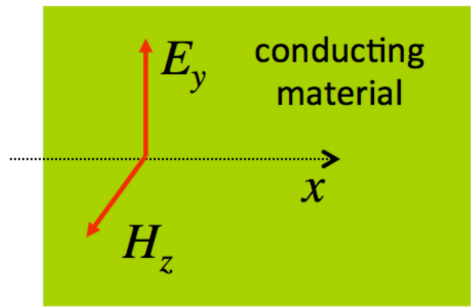

**Fig. A.3:** Plane wave propagating inside a conducting material

From Maxwell's equations, we get the wave equation for the electric field:

$$\dfrac{\partial^2 E_y}{\partial x^2} - \varepsilon\mu \dfrac{\partial^2 E_y}{\partial t^2} - \sigma\mu \dfrac{\partial E_y}{\partial t} = 0.$$

To find the solution of the wave equation, we assume that the electric field propagates in the $x$ direction with as

$$E_y = \tilde{E}_0 e^{i\omega t - \gamma x}.$$

If we substitute the above expression into the wave equation, we obtain the equation for the complex amplitude of the electric field $\tilde{E}_0$:

$$(\gamma^2 + \varepsilon\mu\omega^2 - i\omega\mu\sigma)\tilde{E}_0 e^{i\omega t - \gamma x} = 0.$$

An analogous equation holds for $H_z$. In order to have non-zero electric field, the term inside the parentheses must be zero. If $\sigma \gg \omega\varepsilon$, this reduces to

$$\gamma \cong (1+i)\sqrt{\frac{\sigma\mu\omega}{2}}.$$

Under such a condition we say that the material behaves like a conductor. Since $\gamma$ has a real part, fields propagating in the material are attenuated. The attenuation constant, measured in metres, is called the skin depth:

$$\delta_w \cong \frac{1}{\Re(\gamma)} = \sqrt{\frac{2}{\omega\sigma\mu}}.$$

The skin depth depends on the material properties and the frequency. Copper, for example, has a skin depth

$$\delta_w \cong \frac{6.66}{\sqrt{f}}(\text{cm}).$$

If we assume a beam pipe 2 mm thick, we find that fields up to frequencies of 1 kHz pass through the wall.

Time-varying fields generally pass through the conductor wall if the skin depth is larger than the wall thickness. This happens at relatively low frequency when $\delta_w$ is large, whereas at higher frequencies, for a good conductor, the skin depth is very small, much lower than the wall thickness, so we can assume that both electric and magnetic fields vanish inside the wall. In this situation, the electric field lines are perpendicular to the wall surface, as in the static case, while the magnetic field lines are tangential to the wall. As a consequence, to obtain a time-varying electric field close to a good conductor we can still use the method of the images, while for the magnetic field it is easy to see that we can use the method (shown in Fig. A.2) of changing the direction of the image current.

**Appendix B: Longitudinal forces**

To derive the relationship between the longitudinal and transverse forces inside a beam, let us consider the case of cylindrical symmetry and ultra-relativistic bunches. We know from Faraday's law of induction that a varying magnetic field produces a rotational electric field:

$$\oint \vec{E}\cdot d\vec{l} = -\frac{\partial}{\partial t}\int_S \vec{B}\cdot\hat{n}\, dS.$$

To obtain the longitudinal electric field, we choose as the path for the circulation a rectangle going through the beam pipe (a cylinder of radius $b$) and the beam, parallel to the $z$ axis and with radius $a$, as shown in Fig. B.1. For a generic position $r < a$, and by taking $\Delta z$ small enough so that we can consider the electric field constant, we have

$$E_z(r,z)\Delta z + \int_r^b E_r(r, z+\Delta z)dr - E_z(b,z)\Delta z - \int_r^b E_r(r,z)dr = -\Delta z\frac{\partial}{\partial t}\int_r^b B_\varphi(r)dr.$$

We now write $E_r(r, z + \Delta z) - E_r(r, z) = \dfrac{\partial E_r(r,z)}{\partial z} \Delta z$, so that from the preceding equation we get

$$E_z(r,z) = E_z(b,z) - \int_r^b \left[ \frac{\partial E_r(r,z)}{\partial z} + \frac{\partial B_\varphi(r,z)}{\partial t} \right] dr .$$

By considering the fact that $z = -\beta c t$, we can also write

$$E_z(r,z) = E_z(b,z) - \frac{\partial}{\partial z} \int_r^b \left[ E_r(r,z) - \beta c B_\varphi(r,z) \right] dr .$$

Since the transverse electric field and the azimuthal magnetic field are related by $B_\varphi = \dfrac{\beta}{c} E_r$, we finally obtain

$$E_z(r,z) = E_z(b,z) - (1-\beta^2) \frac{\partial}{\partial z} \int_r^b E_r(r,z) dr .$$

Note that for perfectly conducting walls we have $E_z(b,z) = 0$, so that

$$E_z(r,z) = -\frac{1}{\gamma^2} \frac{\partial}{\partial z} \int_r^b E_r(r,z) dr .$$

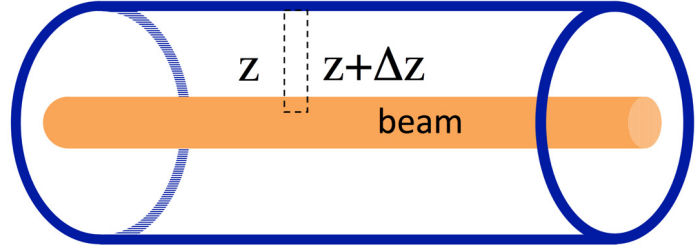

**Fig. B.1:** Geometry for obtaining the longitudinal electric field due to space charge